# Signature of Orbital Driven Finite Momentum Pairing in a 3D Ising Superconductor

F. Z. Yang[1,#], H. D. Zhang[1,#], Saswata Mandal[2,#], F. Y. Meng[3,4,#], G. Fabbris[5], A. Said[5], P. Mercado Lozano[5], A. Rajapitamahuni[6], E. Vescovo[6], C. Nelson[6], S. Lin[1], Y. Park[1], E. M. Clements[1], T. Z. Ward[1], H.-N. Lee[1], H. C. Lei[3,4,*], C. X. Liu[2,*], H. Miao[1,*]

[1]*Materials Science and Technology Division, Oak Ridge National Laboratory, Oak Ridge, Tennessee 37831, USA*

[2]*Department of Physics, The Pennsylvania State University, University Park, PA 16802, USA*

[3]*Department of Physics and Beijing Key Laboratory of Opto-electronic Functional Materials & Micro-nano Devices, Renmin University of China, Beijing 100872, China*

[4]*Key Laboratory of Quantum State Construction and Manipulation (Ministry of Education), Renmin University of China, Beijing 100872, China*

[5]*Advanced Photon Source, Argonne National Laboratory, Argonne, Illinois 60439, USA*

[6]*National Synchrotron Light Source II, Brookhaven National Laboratory, Upton, New York 11973, USA*

**The finite momentum superconducting pairing states (FMPs), where Cooper pairs carry non-zero momentum, are believed to give rise to exotic physical phenomena including the pseudogap phase of cuprate high-$T_c$ superconductors and Majorana fermions in topological superconductivity. FMPs can emerge in intertwined electronic liquids with strong spin-spin interactions or be induced by lifting the spin degeneracy under magnetic field as originally proposed by Fulde-Ferrell and Larkin-Ovchinnikov. In quantum materials with strong Ising-type spin-orbit coupling, such as the 2D transition metal dichalcogenides (TMDs), the spin degree of freedom is frozen enabling novel orbital driven FMPs via magnetoelectric effect. While evidence of orbital driven FMPs has been revealed in bilayer TMDs, its realization in 3D bulk materials remains an unresolved challenge. Here we report experimental signatures of FMP in a locally noncentrosymmetric bulk superconductor 4$H_b$-$TaS_2$. Using hard X-ray diffraction and angle-resolved photoemission spectroscopy, we reveal unusual 2D chiral charge density wave (CDW) and weak interlayer hopping in 4$H_b$-$TaS_2$. Below the superconducting transition temperature, the upper critical field, $H_{c2}$, linearly increases via decreasing temperature, and well exceeds the Pauli limit, thus establishing the dominant orbital pair-breaking mechanism. Remarkably, we discover a field-induced superconductivity-to-superconductivity transition that breaks continuous rotational symmetry of the s-wave uniform pairing in the Bardeen-Cooper-Schrieffer theory down to the six-fold rotation symmetry. Combining with a Ginzburg-Landau free energy**

**analysis that incorporates magnetoelectric effect, our observations provide strong evidence of orbital driven FMP in the 3D quantum heterostructure 4$H_b$-$TaS_2$.**

The finite momentum superconducting paring states (FMPs) and their interplay with the high-$T_c$ and topological superconductivity (SC) represent a forefront of condensed matter physics[1-8]. Unlike conventional Bardeen-Cooper-Schrieffer (BCS) superconductors characterized by spatially uniform order parameters, FMPs feature spatially oscillating pairing amplitude, as illustrated in Fig. 1**a** and 1**b**. Since the FMPs and uniform SC are competing orders, the FMPs are thought to arise when uniform SC is suppressed. This condition can be achieved in strongly correlated electronic liquids, such as the cuprate[9-11], Fe-based[12,13], heavy fermion[14,15] and kagome superconductors[16], where spin, charge, and pairing fields are strongly intertwined. Alternatively, FMPs can be energetically favored in low-dimensional BCS superconductors under external magnetic field[17-24], where the spin Zeeman effect suppresses the uniform SC. While the spin degree of freedom usually plays a key role in microscopic mechanisms of FMPs, in transition metal dichalcogenides (TMDs) with large Ising-type spin-orbital coupling (SOC)[25], the electron spins near the *K* and *K*' valley are pinned to the normal direction of the 2D plane, leading to the significant enhancement of the Pauli limit for the in-plane magnetic fields, $B_{\parallel}$[26-29]. Consequently, the orbital degree of freedom becomes critical for SC under $B_{\parallel}$. Particularly, the magnetoelectric (ME) terms (also known as the Lifshitz invariants[30-32]) that couple magnetic fields to the current operator can exist in the Ginzburg–Landau (GL) formalism when inversion symmetry is locally broken, leading to FMPs such as the helical vortex phase[30-32]. Figure 1c depicts an example of an orbital driven FMP mechanism for 2D bilayer systems[33-36]. However, the experimental signature of orbital driven FMPs in 3D TMDs has not been well established.

The 4$H_b$-$TaS_2$ is an intrinsically correlated TMDs that interweaves Ising SC[28] and charge density wave (CDW) induced flat band (Fig. 1**e**)[37-39]. The crystal structure of 4$H_b$-$TaS_2$ is formed by alternative stacking of the inversion ($\mathcal{P}$)-breaking 1*H*/1*H'*- $TaS_2$ and $\mathcal{P}$-preserving 1*T*-$TaS_2$ layers (Fig. 1**d**). 4$H_b$-$TaS_2$ belongs to a class of locally noncentrosymmetric superconductors[40], in which $\mathcal{P}$ connects 1*H* and 1*H'* layers with Ising SC. Figure 1**f** shows the resistivity of 4$H_b$-$TaS_2$. The large jump at 315 K and the quick drop at 25 K correspond to CDWs in the 1*T* and 1*H*/1*H'* sublayers, respectively[41]. Below $T_{SC}$ = 3 K, 4$H_b$-$TaS_2$ is a clean limit Ising superconductor[42] and displays signatures of spontaneous time-reversal symmetry breaking[43-45]. In this letter, we report the experimental observations of 2D electronic states in the 3D bulk $4H_b$-$TaS_2$ and provide experimental and theoretical evidence of orbital driven FMP via magnetoelectric effects under magnetic fields.

**The 2D electronic structures of the normal state**

Figure 2**a** depicts CDW superlattice peaks in the $1T$ layers. The 2D star-of-David (SoD) lattice distortions breaks the mirror symmetry and result in right-handed (orange) and left-handed (cyan) superlattice peaks that rotate $\pm 13.9°$ with respect to the fundamental Bragg peaks (green)[41]. Figure 2**b** and 2**c** show the intensity distributions of CDW superlattice peaks at $Q^r$ and $Q^l$ in the HL-diffraction plane, respectively. Surprisingly, while the chiral CDW is ordered in the 2D plane, it is completely disordered in the stacking direction resulting in diffraction rods along the L-direction. This observation is in stark contrast to the 3D crystal structure (see Supplementary Materials), establishing emergent 2D electronic state in 3D systems. Figure 2**d**-**f** show the angle-resolved photoemission spectroscopy (ARPES) determined electronic structure of $4H_b$-$TaS_2$ at photon energy, $h\nu$=95 eV, and temperature, $T$ = 30 K. Due to charge transfers between the $1T$ and $1H$/$1H'$ layers, the SoD CDW induced flat band is crossing the Fermi level and forming chiral Fermi surfaces (FSs) around the surface projected $\overline{\Gamma}$ point in $4H_b$-$TaS_2$. In Fig. 2**d**, we reveal the windmill-shaped FS[46] using a beam size of ~1 μm. The cyan hexagon indicates the folded Brillouin zone in the left-handed CDW domains.

The dog-bone-shaped FSs near the $\overline{\mathrm{M}}$ and $\overline{\mathrm{K}}$ points are electronic states from the $1H$/$1H'$-$TaS_2$[47]. Because of the Ising SOC, two electronic bands with opposite spin are split along the $\overline{\Gamma}$-$\overline{\mathrm{K}}$ point direction, as shown in Fig. 2**e**. To quantify the size of SOC in the $1H$/$1H'$ layers, we fit the electronic structure using a tight binding model[48]. Figure 2**f** shows the fitted result on top of the ARPES intensity plot along the $\overline{\mathrm{K}}$-$\overline{\mathrm{M}}$-$\overline{\Gamma}$-$\overline{\mathrm{K}}$ path. The extract $\beta_{soc}$ is about 220 meV. Furthermore, in agreement with the 2D CDW shown in Fig. 2**b** and 2**c,** we find that the electronic structure is non-dispersive along the $k_z$-direction, deriving an upper limit of $\frac{t_\perp}{\beta_{soc}} < 0.02 \ll 1$ in $4H_b$-$TaS_2$ (see Supplementary Fig. S3).

**Emergent quasi-2D superconductivity in $4H_b$-TaS$_2$**

The observations of 2D electronic states with $\frac{t_\perp}{\beta_{SOC}} \ll 1$ indicate that the spin degree of freedom is frozen and hence explain the enhanced upper critical field beyond the standard Pauli limit under $B_\parallel$ in the 3D bulk $4H_b$-TaS$_2$[40]. Figures 3**a** and **b** show the magnetic field dependence of current density ($J$) vs voltage (V) maps, $V(B_\perp, J)$ and $V(B_\parallel, J)$ at $T$ = 30 mK, where $B_\perp$ is the magnetic field perpendicular to the 2D plane. The SC, depicted as the dark blue region, is vulnerable to $B_\perp$ as expected for clean limit BCS superconductors, but robust under $B_\parallel$ as expected for Ising SC. Indeed, the zero-temperature upper critical field, $H_{c2}$, is about 4 times of the standard Pauli limit under $B_\parallel$ according to $B_P$ =1.86$T_{SC}$ ~ 4.2 T (Fig. 3**c**). This is because the Zeeman term of in-plane magnetic field can only couple a spin-polarized band at *K/K'* valley in the 1$H$ layer to the opposite spin polarized band at the same valley in the 1$H'$ layer due to the Ising SOC (Fig. 1**c**). Consequently, the corresponding in-plane Zeeman spin splitting is reduced by a factor of $\frac{t_\perp}{\beta_{SOC}} < 0.02$[33]. Thus, the in-plane upper critical field is mainly determined by the orbital pair-breaking mechanism for the 3D bulk $4H_b$-TaS$_2$.

Figure 3**d** shows the polar angle, $\theta$, dependence of $H_{c2}(\theta)$ at 30 mK. The experimental geometry is shown in the inset of Fig. 3**b**. Tilting magnetic field away from $\theta$ =0°, $H_{c2}(\theta)$ displays humps at $\theta_c \sim \pm 1°$, suggesting a SC-to-SC transition. Fitting of the experimental curve shows that $H_{c2}(\theta > \theta_c)$ is described by the anisotropic 3D GL model, $(H_{c2}(\theta)\cos\theta / H_{c2}^{\parallel})^2 + (H_{c2}(\theta)\sin\theta / H_{c2}^{\perp})^2 = 1$, whereas $H_{c2}(\theta < \theta_c)$ is captured by the 2D Tinkham model, $(H_{c2}(\theta)\cos\theta / H_{c2}^{\parallel})^2 + |H_{c2}(\theta)\sin\theta / H_{c2}^{\perp}| = 1$[49]. We thus conclude that the $H_{c2}(\theta_c)$ anomaly corresponds to a transition from 3D BCS state, where magnetic flux pass through the Abrikosov vortex, as depicted in Fig. 3**e**, to an unconventional quasi-2D SC, where the Josephson vortex is confined in the inversion-symmetric 1$T$-TaS$_2$ layers, as depicted in Fig. 3**f**. These observations establish that the superconducting phase of $4H_b$-TaS$_2$ under large $B_\parallel$ is an effective 2D bilayer system, where the 1$H$ and 1$H'$ layers are weakly coupled via Josephson tunneling assisted by the electronic states at 1$T$ layers.

**Rotational invariance breakdown and FMP**

To determine the nature of the 2D SC under magnetic field, we determine the magnetoresistance, $R(B_{\parallel}, \varphi)$ of $4H_b$-$TaS_2$, where $\varphi$ is azimuthal angle defined in the inset of Fig. 3**b**. Figure 4**a** shows the polar plot of $R(B_{\parallel}, \varphi)$ at $T$=1.9 K. We find that under small $B_{\parallel}(\varphi)$, the $R(B_{\parallel}, \varphi)$ is twofold symmetric. This observation is consistent with a spatially isotropic superconductivity, where the continuous rotational symmetry $C_{\infty}$ is reduced to $C_2$ in the presence of external currents. Indeed, we have confirmed that the $C_2$ axis is always pinned to the current direction in all measurements with different samples and geometries (see Supplementary Materials). Remarkably, increasing $B_{\parallel}(\varphi)$ at a fixed temperature to a critical $B_c$, we find a $C_2$ to $C_6$ transition. Figure 4**b** compares the two-fold symmetric $R(B_{\parallel} = 1T, \varphi)$ and the sixfold symmetric $R(B_{\parallel} = 4T, \varphi)$. The $R(B_{\parallel} = 1T, \varphi)$ exhibits a maxima and minima at $B \perp I$ and $B // I$ configurations, respectively. The $R(B_{\parallel} = 4T, \varphi)$ shows minima when magnetic field directions are parallel or anti-parallel to the crystalline direction at $\varphi$ = -60°, 0° and 60°. These observations establish that the rotational invariance of the isotropic s-wave SC is broken down to six-fold rotation $C_6$ that is determined by the crystal directions under high $B_{\parallel}(\varphi)$. Increasing temperature above $T_{SC}$, as shown in Fig. 4**d**, restores the $C_2$ symmetry as expected for the isotropic Fermi liquid normal state. The recovery of $C_2$ in the normal state confirms that the $C_6$ symmetric SC under magnetic field is an intrinsic 2D superconducting phase that is different from the spatially isotropic SC. Similar to finite-$q$ electronic orders, such as the spin and charge density waves, the FMPs will also break the continuous rotational symmetry as a consequence of directional translational symmetry breaking. Therefore, the observations of SC-SC transition and rotational invariance breakdown provide experimental evidence of an orbital-driven FMP in $4H_b$-$TaS_2$. Indeed, a similar rotational symmetry breaking of SC pairing was also reported in $NbSe_2$ multilayers[35] and was interpreted as the signature of orbital driven FMPs.

To understand the orbital driven FMPs in 3D, we consider a Lawrence-Doniach type of model[50,51] with the standard superconducting GL free energy for each $1H$/$1H'$-$TaS_2$ layer. The adjacent superconducting layers are coupled by a weak Josephson tunneling and the orbital effect of in-plane magnetic fields is incorporated via replacing the spatial derivative by covariant derivative (see Methods). Additional ME coupling terms are allowed in the GL free energy due to the local inversion symmetry breaking around the $1H$/$1H'$-$TaS_2$ layer even though the global inversion exists[40]. The mirror symmetries of the $1H$/$1H'$-$TaS_2$ layers are further broken down, as evidenced by the observations of chiral CDW and forbidden Bragg peaks (See Supplementary Materials). Our numerical simulations show that the ME coupling can give rise to the FMP in the intermediate magnetic field range (See Supplementary Materials). Importantly, the higher-order terms of ME coupling naturally lead to the breaking of rotational invariance of SC pairing down to six-fold rotation $C_6$, in agreement with the experimental observations (See Supplementary Materials).

Figure 4**e** summarizes the experimentally determined $B_{\parallel}$-$T$ phase diagram. The extracted phase boundaries between BCS, FMPs, and Fermi liquid normal states are marked by black and red dots. When the $B_{\parallel}$ is lower than $B_c$, the system is a uniform Josephson coupled Ising superconductor. When $B_{\parallel}$ is greater than $B_c$, the FMPs state is formed in 1*H* and 1*H'* layers. The proximity induced SC in 1*T* layer[44] is completely suppressed by external magnetic fields due to the absence of Ising-type SOC and the formation of Josephson vortices. Physically, the first order phase boundary separating the uniform SC and FMPs is determined by competition between interlayer Josephson coupling and the energy gain of FMP under in-plane magnetic fields. Due to the dominating orbital effect, the $H_{c2}$ separating the normal metal and FMP is linear in *T*, which is also captured by our GL calculations (Supplementary Materials).

Finally, we discuss the multi-component SC scenarios proposed for $4H_b$-$TaS_2$[52,53]. First, we note that SC transition temperatures of $4H_b$-$TaS_2$ and monolayer $1H$-$TaS_2$ are close to each other, suggesting the same s-wave Ising SC pairing mechanism in both systems. Nevertheless, as we showed in Fig. 2, the chiral CDW induced flat band in the 1*T* layer is crossing the Fermi level, which can host a uniform chiral SC in the 1*T* layer via the superconducting proximity effect. However, the rotational-invariance breakdown at high $B_{\parallel}$ is unlikely related to SC in the 1*T* layer due to the absence of Ising SOC and small superfluid density in the 1T-layer.

In summary, we uncovered unusual 2D electronic states in a 3D quantum heterostructure. The magnetic field induced SC-to-SC transition supports the emergence of orbital driven FMPs in $4H_b$-$TaS_2$. Our discovery opens new avenue to realize unconventional SC and highlight the bulk TMDs quantum heterostructures as ideal platform for novel quantum states.

## Methods

**Sample preparation and characterizations:**

High-quality single crystals of $4H_b$-$TaS_2$ were grown by using the chemical vapor transport method[54]. A stoichiometric mix of Ta and S powders with additional 0.15 g of $I_2$ was sealed under high vacuum in silicon quartz tubes. These tubes heated for 15 days in a two-zone furnace, where the temperature of source and growth zones were fixed at 820 ℃ and 750 ℃, respectively.

**X-ray scattering measurements:**

The single crystal elastic X-ray diffraction was performed at the 4-ID-D beamline of the Advanced Photon Source (APS), Argonne National Laboratory (ANL), and the integrated *in situ* and resonant hard X-ray studies (4-ID) beam line of National Synchrotron Light Source II (NSLS-II). The photon energy, which is selected by a cryogenically cooled Si(111) double-crystal monochromator, is 9.88 keV.

**The 4-ID-D, APS**: the X-rays higher harmonics were suppressed using a Si mirror and by detuning the Si (111) monochromator. Diffraction was measured using a vertical scattering plane geometry and horizontally polarized (σ) X-rays. The incident intensity was monitored by a $N_2$ filled ion chamber, while diffraction was collected using a Si-drift energy dispersive detector with approximately 200 eV energy resolution. The sample temperature was controlled using a He closed cycle cryostat and oriented such that X-rays scattered from the (001) surface.

**The 4-ID, NSLS2**: The sample is mounted in a closed-cycle displex cryostat in a vertical scattering geometry. The incident X-rays were horizontally polarized, and the diffraction was measured using a silicon drift detector.

**ARPES measurements:**

The ARPES experiments are performed at beamline 21-ID-1 of NSLS-II at BNL. The $4H_b$-$TaS_2$ samples are cleaved in situ in a vacuum better than $3 \times 10^{-11}$ Torr. The measurements are taken with synchrotron light source and a Scienta-Omicron DA30 electron analyzer with a beam size ~ 1 μm. The total energy resolution of ARPES measurement is approximately 15 meV. The sample stage is maintained at 30 K throughout the experiments.

**Transport measurements:**

Transport measurements were performed on Triton Cryofree Dilution Refrigerator (Oxford instruments) and Physical Property Measurement System (Quantum Design). Polar angle ($\theta$) dependent resistivity data were measured on a piezo rotator with built-in resistive gauge. We use a lock-in amplifier (SR860, Stanford Research) for $\theta$ readings, with typical angular resolution at about 0.03 degrees. The resistivity measurement current is supplied by K6221 source meter (Keithley), and the voltage readings were monitored by K2182 nanovolt meter (Keithley). Azimuthal angle ($\varphi$) dependent resistivity data were measured on the rotator option from Quantum design. The typical angular resolution is near 1 degree, and one must pay special attention to the backlash problem of the factory default setup. The resistivity data is measured by the resistivity option from Quantum Design.

**Fitting method for the phase diagram:**

The experimental data show both six-fold and two-fold components. We use function, $\cos^2(\varphi)$, to fit the twofold background. After subtraction, we define the amplitude of six-fold component $R_{six}$ as the average of resistance value at $\varphi = 0, \pm 60$ deg. At a fixed temperature, the $R_{six}$ displays a linear dependence on the magnetic field $B_{\parallel} > B_c$. Extrapolating the linear line to $R_{six} = 0$ gives the phase boundary between the uniform SC and FMP states. The error bars of the phase boundary represent the standard deviations of the linear fittings.

**Ginzburg-Landau Theory of bulk layered Ising superconductors:**

To model the bulk $4H_b$-$TaS_2$, we consider the Lawrence-Doniach type of model[50,51,34] in the context of GL theory with the form $\mathcal{F} = \mathcal{F}_0 + \mathcal{F}_J + \mathcal{F}_{ME}$, where

$$\mathcal{F}_0 = \sum_{l\eta} \int d^2r \left( \alpha \left|\psi_{l,\eta}\right|^2 + \frac{\beta}{2}\left|\psi_{l,\eta}\right|^4 + \frac{\hbar^2}{2m^*}\left|\vec{D}_{l\eta}\psi_{l,\eta}\right|^2 \right)$$

$$\mathcal{F}_J = -J_0 \sum_{l} \int d^2r \left( \psi_{l,+}\psi_{l,-}^* + \psi_{l+1,-}\psi_{l,+}^* + h.c. \right)$$

$$\mathcal{F}_{ME} = \sum_{l\eta} \eta\gamma_0 \int d^2r \left( B_y j_{l,\eta}^x - B_x j_{l,\eta}^y \right).$$

Here we consider the $TaS_2$ layers stacked along the z direction and $\boldsymbol{r} = (x, y)$ for the in-plane directions. $\psi_{l,\eta}$ is the superconducting order parameter with $\eta = \pm$ the layer index for $1H/1H'$ layer in the unit cell $l$. $J_0$ is the Josephson interlayer coupling parameter, $\gamma_0$ is the strength of the ME effect, $\alpha, \beta$ are the coefficients of the second and fourth order terms in order parameter. We choose $\alpha = \alpha_0(T - T_0)$ with the parameter $\alpha_0$ and $T_0$. The covariant derivative $\vec{D}_{l,\eta} = -\mathrm{i}\,\vec{\nabla} + \frac{2e}{\hbar}\vec{A}_{l,\eta}$ and the current operator is given by $j_{l,\eta}^i = \psi_{l,\eta}^* D_{l,\eta}^x \psi_{l,\eta} + \psi_{l,\eta}\left(D_{l,\eta}^x \psi_{l,\eta}\right)^*$. $\mathcal{F}_0$ models the superconducting state in $1H/1H'$-$TaS_2$ layers, $\mathcal{F}_J$ describes the inter-layer Josephson coupling, and $\mathcal{F}_{ME}$ is the ME coupling term[30-32] due to the local inversion symmetry breaking (see Supplementary Materials). Here we assume the inter-layer Josephson coupling is weak, and the parameter $J_0$ can be positive or negative, depending on the inter-layer electron hopping form. Particularly, it was suggested[54] that the negative Josephson coupling $J_0 < 0$, together with dislocation, can provide an explanation of the $\pi$ phase shift of the Little-Parks experiments in the $4H_b$-$TaS_2$[53]. The full form of the ME coupling term $\mathcal{F}_{ME}$ can be constructed from the crystal symmetry in the Supplementary Materials and the form of $\mathcal{F}_{ME}$ shown above is a simplified version and used for our numerical calculations. Here we consider the $C_{3v}$ point group symmetry for the $1H/1H'$-$TaS_2$ layers, in which the mirror symmetry along the *z* direction is broken as supported by the x-ray diffraction (see Supplementary Materials). We also note that the in-plane magnetic field $\vec{B} = \left(B_x, B_y\right)$ appears in the covariant derivative $\vec{D}_{l,\eta}$ and the ME coupling term $\mathcal{F}_{ME}$. Both $\vec{D}_{l,\eta}$ and $\mathcal{F}_{ME}$ are connected to the orbital effect of magnetic fields. The Zeeman effect of magnetic fields is dropped in the above GL free energy. This is because the Ising SOC suppresses the Zeeman effect and significantly enhances the Pauli limit[26-29]. Thus, the above GL free energy $\mathcal{F}$ is suitable for the description of the bulk layered Ising superconducting materials.

For the in-plane magnetic field $\vec{B} = \left(B_x, B_y\right)$, we can choose the vector potential as $\vec{A}_{l\eta} = \left(2l + \frac{\eta}{2}\right) d_0\left(\hat{z} \times \vec{B}\right)$, where $d_0$ is the distance between two adjacent $1H$ and $1H'$ layers, so the lattice constant along the *z* direction is $2d_0$. As the vector potential $\vec{A}_{l\eta}$ relies on the unit cell $l$, the translation symmetry along the *z* direction is broken. We perform a gauge transformation

$$\tilde{\psi}_{l,\eta}(\vec{r}) = e^{-i\vec{q}_{l\eta}\cdot r}\psi_{l,\eta}(\vec{r}),$$

With $\vec{q}_{l,\eta} = \frac{2e}{\hbar}\left(2l + \frac{\eta}{2}\right) d_0\left(\hat{z} \times \vec{B}\right)$, to restore the *z*-directional translation invariance with the price that the in-plane continuous translation symmetry is broken down to the discrete translation by forming the magnetic unit cell. We then carry out the Fourier transformation for $\tilde{\psi}_{l,\eta}(\vec{r})$ into the momentum space and obtain

$$\tilde{\psi}_{l,\eta}(\vec{r}) = \frac{1}{\sqrt{N_z S}} \sum_{n,\vec{k}} e^{il\frac{2\pi n}{N_z}} e^{i\vec{k}\cdot\vec{r}} \psi_{n\eta}(\vec{k})$$

where $N_z$ is the total number of unit cells and $S$ defines the area of each layer. After the Fourier transformation, the GL free energy in the momentum space takes the form

$$\mathcal{F} = \sum_{n,\vec{k},\eta} \left( \alpha + \frac{\hbar^2 k^2}{2m^*} + 2\eta\gamma_0 (k_x B_y - k_y B_x) \right) \left|\psi_{n,\eta}(\vec{k})\right|^2$$
$$- J_0 \sum_{nk} \psi_{n+}(\vec{k}) \psi^*_{n-}(\vec{k} + \delta\vec{q}) + e^{-2\pi n/N_z} \psi_{n+}(\vec{k}) \psi^*_{n-}(\vec{k} - \delta\vec{q}) + c.c$$
$$+ \frac{\beta}{2S} \sum_{n,n',m} \sum_{k_1,k_2,p} \psi^*_{n\eta}(k_1) \psi^*_{m\eta}(k_2) \psi_{n'\eta}(k_1 - p) \psi_{n+m-n'\eta}(k_2 + p).$$

The GL equation can be derived by taking the functional derivative $\frac{\partial \mathcal{F}}{\partial \psi^*_{l\eta}(k)} = 0$, and has the form

$$-\left\{ \frac{\hbar^2 k^2}{2m^*} + 2\eta\gamma_0 (k_x B_y - k_y B_x) \right\} \psi_{n\eta}(\vec{k}) + J_0 \delta_{\eta+} \left[ \psi_{n-}(\vec{k} + \delta\vec{q}) + e^{i2\pi n/N_z} \psi_{n-}(\vec{k} - \delta\vec{q}) \right]$$
$$+ J_0 \delta_{\eta-} \left[ \psi_{n+}(\vec{k} - \delta\vec{q}) + e^{-i2\pi n/N_z} \psi_{n+}(\vec{k} + \delta\vec{q}) \right] = \alpha_0 (T - T_0) \psi_{n\eta}(\vec{k}).$$

Here we have linearized the GL equation by dropping the fourth order terms, which is a valid approximation for the superconducting states close to the critical temperature so that the SC order parameter is a small number. The above GL equation has a similar form as the Bloch equation for electrons in a periodic potential, and the momentum shift $\delta\vec{q}$ plays the role of reciprocal lattice vector. This GL equation can be regarded as an eigen-equation problem, which can be solved numerically (See Supplementary Materials), and the eigen-values give $\alpha_0(T - T_0)$, which is a function of momentum $\vec{k} = (k_x, k_y)$ and $k_z = \frac{2\pi n}{N_z}$. The largest eigen-value for $T(\vec{k}, n)$ gives rise to the SC critical temperature $T_c$ in the channel $(\vec{k}, n)$. We maximize $T(\vec{k}, n)$ with respect to both $\vec{k}$ and $n$ to get the true $T_c$ for the system and if the corresponding optimal momentum, denoted as $\vec{k}_{\mathbf{0}}$, is non-zero, we obtain the superconducting state with FMP. Our numerical results suggest the FMPs can exist in certain range of in-plane magnetic fields as a result of non-zero ME coupling term $\gamma_0$. To see how ME coupling term $\gamma_0$ can induce FMP, we may consider the limit with zero inter-layer Josephson coupling $J_0 = 0$, which essentially corresponds to the 2D limit, the critical temperature for the momentum channel $\vec{k}$ is then $T(\vec{k}) = T_0 - \frac{1}{\alpha_0}\left(\frac{\hbar^2 k^2}{2m^*} + 2\eta\gamma_0(k_x B_y - k_y B_x)\right) = T_0 - \frac{\hbar^2}{2m^*}\left(\vec{k} + \eta\gamma\frac{2m^*}{\hbar^2}(\vec{B} \times \hat{z})\right)^2 + \frac{2m^*\gamma^2}{\hbar^2}B^2$. Thus, in this limit the critical temperature is given by $T_c = T_0 + \frac{2m^*\gamma^2}{\hbar^2}B^2$ for a finite momentum $\vec{k}_0 = -\eta\gamma\frac{2m^*}{\hbar^2}(\vec{B} \times \hat{z})$. More detailed behaviors of the superconducting states with FMPs for the general choice of parameters are discussed in the Supplementary Materials.

We also notice that the crystal symmetry also allows for higher-order ME coupling terms in the free energy

$$\mathcal{F}^{(1)}_{ME} = \sum_{n,\vec{k}} \Omega^\dagger_n(\vec{k}) \{ \gamma_1 Im[k_\pm B^2_\mp] \tau_2 + \gamma_2 Re\left[k^2_\pm B^4_\mp\right] \tau_0 + \gamma_3 Re\left[k^4_\pm B^2_\mp\right] \tau_0 \} \Omega_n(\vec{k})$$

where $\Omega_n(\vec{k}) = \{\psi_{n+}(\vec{k}), \psi_{n-}(\vec{k})\}^T$ is the eigenstate for a bilayer unit cell and $\tau$ acts in the layer basis ($\eta = \pm$) in each unit cell. We treat this term as a perturbation and its correction to the critical temperature is derived as

$$\Delta T_c(B,\phi) = \Psi_0^\dagger(k_0)\mathcal{F}_{ME}^{(1)}\Psi_0(k_0)$$

where $\vec{B} = B(\cos\phi, \sin\phi)$ and the $k_0$ is the optimal momentum found from the lowest order G-L equation, which can be solved numerically for the critical temperature, and $\Psi_0(k_0)$ is the eigenstate corresponding to this solution. Our numerical results in Fig. S13 of Supplementary Materials show the breaking of full rotation symmetry to six-fold rotation symmetry of $T_c$ due to the higher-order ME coupling term $\mathcal{F}_{ME}^{(1)}$.

**Data availability**

The data that support the findings of this study are available from the corresponding author upon reasonable request.

**Acknowledgements**

We thank Andrew Christianson, Gang Chen, Hong Ding, Chenyun Hua, Lingyuan Kong, Patrick A. Lee, Wenyao Liu, Qiangsheng Lu, Andrew May, Michael McGuire, Thomas Maier, John Tranquada, Ziqiang Wang, Binghai Yan, Jiaqiang Yan, Noah F. Q. Yuan, Ruixing Zhang, and Yang Zhang for stimulating discussions. This research was supported by the U.S. Department of Energy, Office of Science, Basic Energy Sciences, Materials Sciences and Engineering Division (X-ray, ARPES, and transport measurements). X-ray scattering used resources (beamline 4ID and 30ID) of the Advanced Photon Source, a U.S. DOE Office of Science User Facility operated for the DOE Office of Science by Argonne National Laboratory under Contract No. DE-AC02-06CH11357. ARPES and X-ray scattering measurements used resources at 21-ID-1, 4-ID and 10ID beamlines of the National Synchrotron Light Source II, a US Department of Energy Office of Science User Facility operated for the DOE Office of Science by Brookhaven National Laboratory under contract no. DE-SC0012704. H.C.L. was supported by Beijing Natural Science Foundation (Grant No. Z200005), National Key R&D Program of China (Grants No. 2022YFA1403800, No. 2023YFA1406500), National Natural Science Foundation of China (Grants No. 12274459). C.X.L. acknowledge the support from the NSF through The Pennsylvania State University Materials Research Science and Engineering Center [DMR-2011839].

**Author contributions**

F.Y., H.D.Z., S.M., and F.Y.M contributed equally to this work. H. M. conceive and designed the research. F.Y., G.F., A.S., P.L., C.N., S.L., Y.P., H.N.L. and H.M. performed X-ray scattering measurements. F.Y., A.R., E.V. and H.M. carried out the ARPES measurements. H.Z., F.Y., and E.M.C. performed the transport measurements. S.M. and C.X.L. performed the theoretical calculations and analysis. F.Y.M. and H.C.L. synthesized the high-quality single crystals. F.Y., H.Z. and H.M. analyzed the experimental data. F.Y., H.Z., H.C.L., C.X.L., and H.M. prepared the manuscript with inputs from all authors.

**Competing interests**

The authors declare no competing interests.

**Additional information**

**Supplementary information** is available at the online version of the paper.

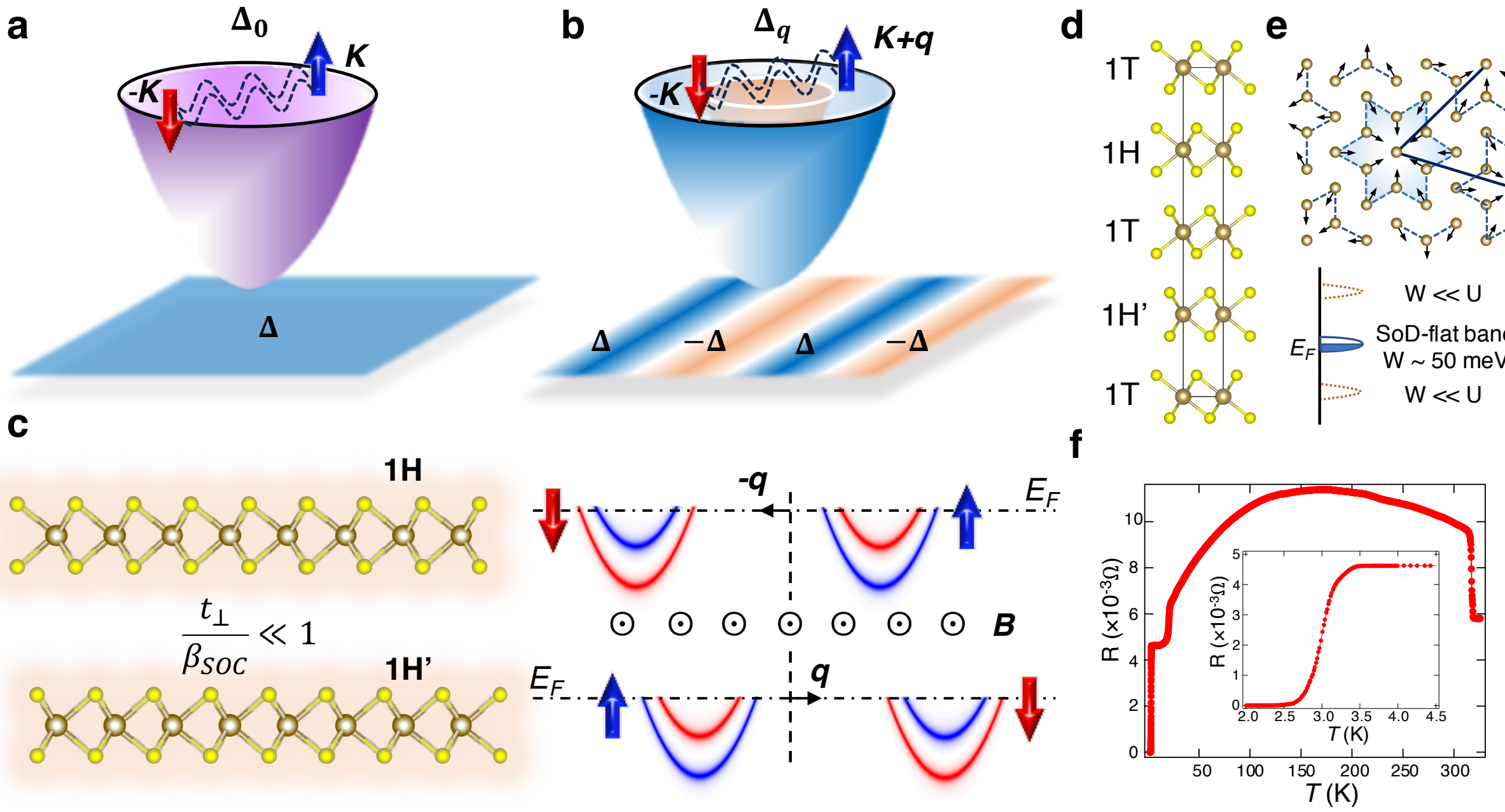

**Figure 1**: **FMPs and orbital effect in TMDs with strong Ising SOC**. **a**, Schematic of conventional BCS superconductivity that has spatially uniform order parameter, $\Delta_0$. Two electrons on the FS with opposite momentum ***K*** and -***K*** form a Cooper pair carrying zero total momentum, $q$=0. **b**, depicts FMPs featuring finite $q$ pairing, $\Delta_q$, and spatially oscillating pairing amplitude. **c**, an orbital driven FMP proposed for locally non-centrosymmetric bilayer TMDs. Despite the global inversion symmetry between 1*H* and 1*H'* layer, when $\frac{t_\perp}{\beta_{SOC}} \ll 1$, the superconducting behavior displays properties of the inversion breaking 1*H*/1*H'* layer, such as the Ising superconductivity. Under external magnetic field shown in **c**, the electronic band structure of 1*H* and 1*H'* layers have opposite momentum shifts, $q_x = \pm\frac{eB_y z_0}{2\hbar}$, due to the opposite gauge field effect or magnetoelectric effect[25]. It is important to note that extension of this 2D model to quasi-2D and 3D quantum materials is yet to be established. **d**, illustration of the crystal structure of 4$H_b$-$TaS_2$. **e**, The star-of-David superlattice and schematic of flat band and Mottness in the 1*T* layer. **f,** Resistivity shows three phase transitions below 350 K. The inset of **f** shows a prototypical resistivity curve near the superconducting transition temperature.

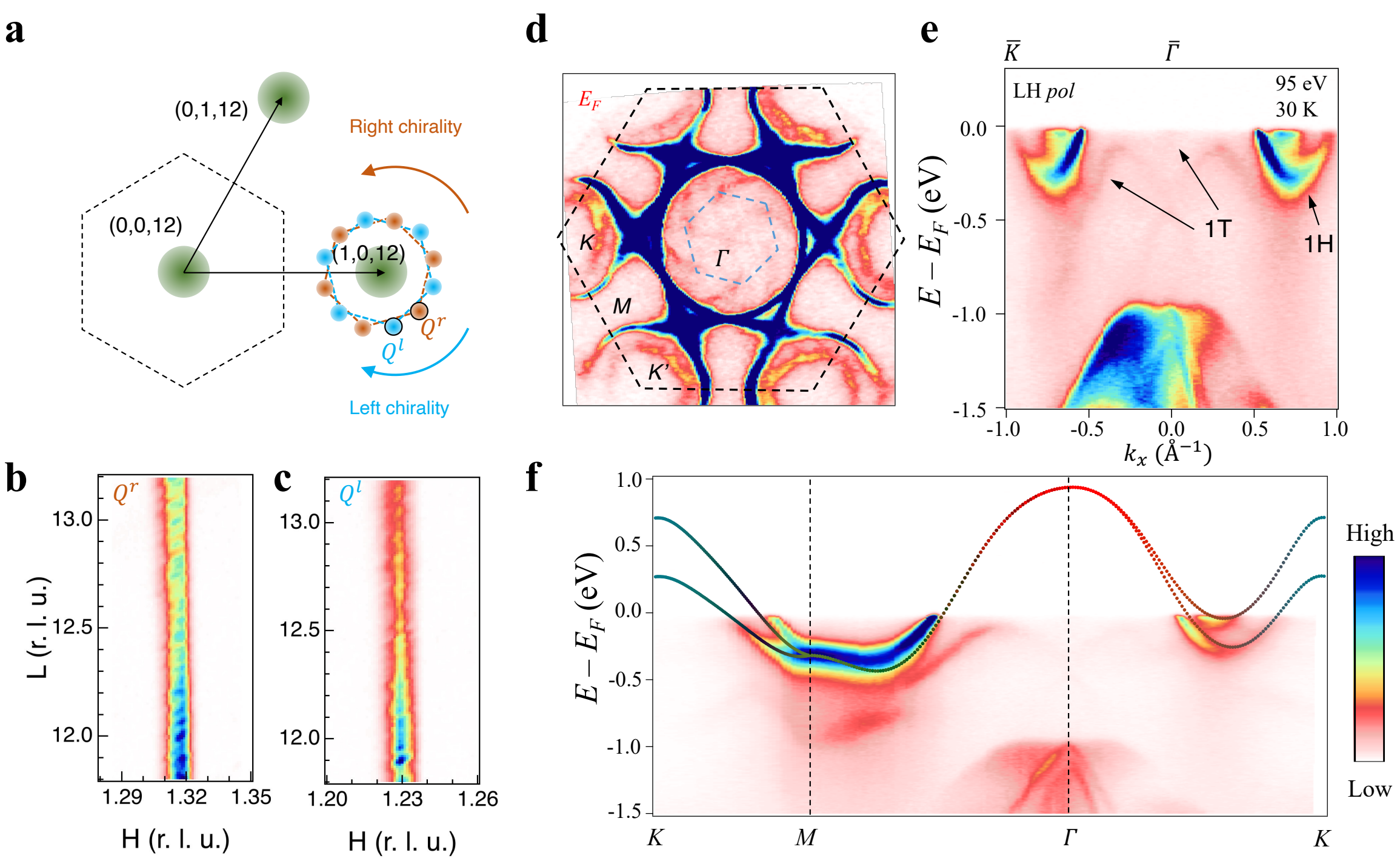

**Figure 2: The 2D electronic structure in the normal state**. **a**, The CDW in the 1*T* layer breaks all mirror symmetries of the 2D plane and induces left (cyan) and right (orange) handed CDW superlattice peaks near the fundamental Bragg peaks (green) in 4$H_b$-$TaS_2$. **b** and **c**, Intensity maps of CDW superlattice peaks, $Q^r$ and $Q^l$, in the HL-scattering plane. The diffraction rods demonstrate a novel 2D CDW in 3D lattice structure. **d**, ARPES determined FS topology of 4$H_b$-$TaS_2$ at 30 K. The black and cyan dashed lines represent the original Brillouin zone and the folded Brillouin zone in the left-handed CDW domain, respectively. The windmill-like FS around the $\overline{\Gamma}$ point arises from the chiral CDW induced flatband. The dog-bone-shaped Fermi surfaces around $\overline{\text{K}}$ and $\overline{\text{M}}$ are from 1*H*/1*H'* layers. **e**, ARPES intensity plot along the $\overline{\text{K}}$-$\overline{\Gamma}$-$\overline{\text{K}}$ path shows both electronic bands of 1*H* and 1*T* layer. **f**, ARPES intensity plot and tight binding fit of the 1*H* layer band structure along the $\overline{\text{K}}$-$\overline{\text{M}}$-$\overline{\Gamma}$-$\overline{\text{K}}$ path.

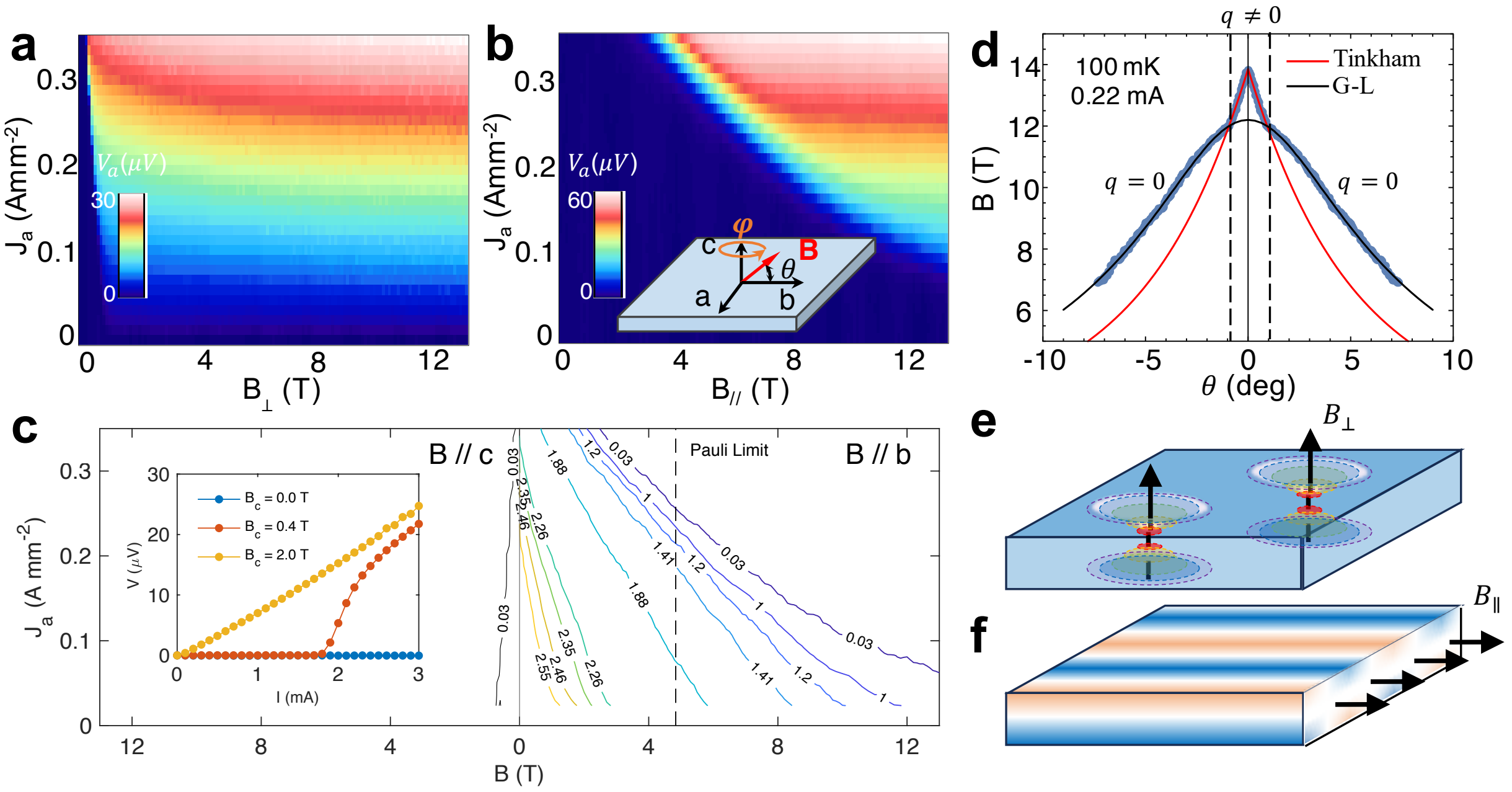

**Figure 3: Polar magnetic field induced 3D SC to 2D SC transition in 4*H*$_b$-TaS$_2$. a and b**, Color maps of $V(B_\perp, J)$ and $V(B_\parallel, J)$ at $T$ = 30 mK, respectively. Dark blue region represents the superconducting phase. Inset in **b** shows the experimental setup. The $\theta$ represents the polar angle, where $\theta$ =0 and 90° are corresponding to $B_\parallel$ and $B_\perp$, respectively. **c**, Phase separation lines between the superconducting state and normal state at various temperatures. The text denotes the measurement temperature in units of Kelvin. Inset: Typical I-V curves data at 30 mK under $B_\perp$=0, 0.4 and 2T. **d**, The polar angle dependence of $H_{c2}(\theta)$. Here $H_{c2}$ is defined as the magnetic field where resistivity reaches 50% of its normal state value. Black and red curves are fittings with the anisotropic Ginzburg-Landau model and 2D Tinkham model, respectively. A SC-to-SC transition around $\theta_c \sim 1°$ is observed. **e** and **f**, Schematics of Abrikosov vortex lattice under $B_\perp$ and Josephson vortex under $B_\parallel$ in 4*H*$_b$-TaS$_2$. Due to the Ising superconductivity in the $\mathcal{P}$-breaking 1*H*-TaS$_2$, Josephson flux is confined in the $\mathcal{P}$-preserving 1*T*-TaS$_2$.

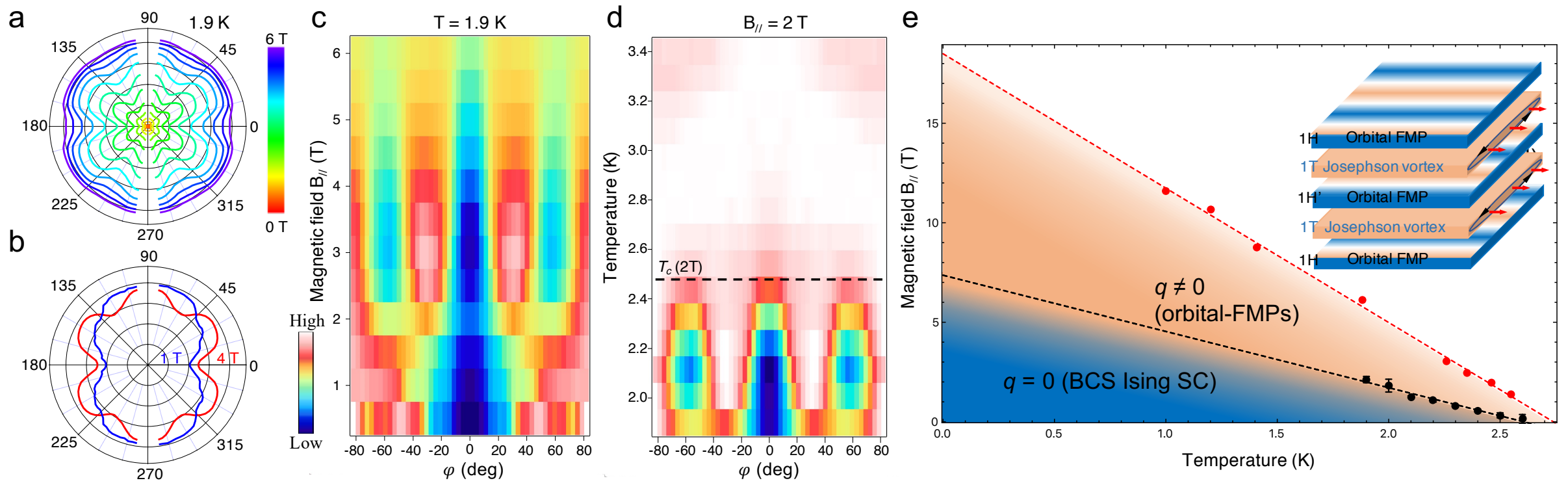

**Figure 4: Signature of orbital driven FMP in 4*H*$_b$-TaS$_2$. a**, Magnetoresistance $R(B_{\parallel}, \varphi)$ at $T =$ 1.9 K. **b**, compares $R(B_{\parallel} = 1T, \varphi)$ and $R(B_{\parallel} = 4T, \varphi)$ showing emergence of $C_6$ symmetry at high field. **c**, Intensity plot of $R(B_{\parallel}, \varphi)$ at T = 1.9 K shows a first order-like $C_2$ to $C_6$ transition at $B_c$. **d**, Intensity plot of the temperature dependent $R(B_{\parallel} = 2T, \varphi)$. The $C_6$ changes back to $C_2$ above the superconducting transition temperature $T_c(2T)$~2.5 K. **e**, The $B_{\parallel}$-$T$ phase diagram of 4$H_b$-TaS$_2$. Experimentally determined phase boundaries between the uniform SC, FMP, and the paramagnetic normal state are marked as black and red dots. The error bars for the $C_2$ to $C_6$ transition (black) represent standard deviations of fitting procedure (see Methods).